\documentclass[
superscriptaddress,
reprint,
nobibnotes,
amsmath,amssymb,
aps,
prl,
floatfix,
longbibliography,
twocolumn
]{revtex4-2}

\usepackage{mathrsfs}
\usepackage{graphicx}
\usepackage{dcolumn}
\usepackage{bm}
\usepackage{chemformula}
\usepackage{braket}
\usepackage{multirow}
\usepackage{booktabs}  
\usepackage{tabularx}  
\usepackage{adjustbox}  
\usepackage{hyperref}

\hypersetup{
    colorlinks=true,
    linkcolor=blue,
    citecolor=blue,
    urlcolor=blue,
}

\usepackage{xr}

\begin{document}

\preprint{APS/123-QED}

\title{First-Principles Electronic Structure Calculation of Crystals in Laboratory Magnetic Fields}

\author{Sichao Wang}
\thanks{These authors contributed equally to this work.}
\affiliation{Key Laboratory of Computational Physical Sciences (Ministry of Education), Institute of Computational Physical Sciences, State Key Laboratory of Surface Physics and Department of Physics, Fudan University, Shanghai 200433, China}

\author{Chengye L\"u}
\thanks{These authors contributed equally to this work.}
\affiliation{Key Laboratory of Computational Physical Sciences (Ministry of Education), Institute of Computational Physical Sciences, State Key Laboratory of Surface Physics and Department of Physics, Fudan University, Shanghai 200433, China}

\author{Xingao Gong}
\affiliation{Key Laboratory of Computational Physical Sciences (Ministry of Education), Institute of Computational Physical Sciences, State Key Laboratory of Surface Physics and Department of Physics, Fudan University, Shanghai 200433, China}

\author{Yingwei Chen}
\email{ywchen17@fudan.edu.cn}
\affiliation{Key Laboratory of Computational Physical Sciences (Ministry of Education), Institute of Computational Physical Sciences, State Key Laboratory of Surface Physics and Department of Physics, Fudan University, Shanghai 200433, China}

\author{Hongjun Xiang}
\email{hxiang@fudan.edu.cn}
\affiliation{Key Laboratory of Computational Physical Sciences (Ministry of Education), Institute of Computational Physical Sciences, State Key Laboratory of Surface Physics and Department of Physics, Fudan University, Shanghai 200433, China}

\date{\today}

\begin{abstract}
External magnetic fields can qualitatively reshape the electronic structure of crystals, underpinning quantum Hall physics, Landau-level spectra and field-induced topological phases. 
Their first-principles treatment at laboratory-scale fields is, however, hindered by magnetic-flux quantization, which requires magnetic unit cells with areas inversely proportional to the applied field. 
Such cells contain a large number of chemical unit cells, rendering real-space and plane-wave calculations prohibitively expensive. 
Here we, for the first time, construct a magnetic Bloch basis built from linear combinations of gauge-including Gaussian-type atomic orbitals, which incorporate the magnetic-field phase factors required by magnetic translation symmetry. 
The framework requires far fewer basis functions than real-space or plane-wave representations of the same magnetic supercell and retains the sparsity of an atom-centred basis, together substantially reducing computational cost. 
We validate the framework by reproducing Landau-level spectrum of graphene from first principles. This approach provides a practical route to simulations of crystalline materials under experimentally accessible magnetic fields.
\end{abstract}

\maketitle

\textit{\label{sec:level1}Introduction}

The behavior of solid materials under an external magnetic field is a topic of great interest in condensed matter physics and materials science.
Magnetic fields exhibit both spin and orbital effects. The spin effect is more direct, while the relatively complex orbital effect is the source of many physical phenomena, such as Landau levels, the Quantum Hall effect\cite{PhysRevLett.49.405}, 
the fractional quantum Hall effect\cite{PhysRevLett.48.1559}, and  strong-field phase transitions\cite{Higuchi_2022}.

As a widely used method in first-principles calculations, the Kohn-Sham density functional theory method \cite{KS1,KS2} can handle the effect of magnetic fields on spin\cite{UvonBarth1972,PhysRevLett.106.107202} well, 
but the calculation method for the orbital effect of magnetic fields is still under development and improvement. For isolated systems (such as molecules), 
calculations using London orbitals carrying field-dependent phase factors have yielded good results\cite{London1,10.1063/1.4861427}; however, for periodic systems, ordinary translational symmetry is replaced by magnetic translation symmetry.

Early researchers mainly used perturbation theory\cite{PhysRevB.56.1009,PhysRevLett.76.4246} or tight-binding methods based on Peierls phase\cite{Peierls0,MWilkinson1984,Peierls2,gll1,PhysRevB.104.035419,PGHarper1955,RevModPhys.81.109,PhysRevLett.126.056401} to simulate the orbital effect of magnetic fields in periodic systems. However, due to limitations imposed by the perturbation order and the applicability of Peierls substitution,
these methods are limited either to perturbative fields or to model-dependent Hamiltonians\cite{PhysRevA.90.033609,PhysRevLett.105.215303,GünterWunner1987}.
In 2004, Cai used a set of ``plane wave-like'' basis sets to characterize magnetic translational symmetry, thus achieving diagonalization of the Hamiltonian under finite magnetic fields. However, his approach was dependent on the size of the basis set and the magnitude of the magnetic field, and it was not suitable for calculating pseudopotentials\cite{PhysRevLett.92.186402,LEE20071310}. 
Pickard introduced magnetic translational symmetry into nonlocal pseudopotentials, proposing the gauge-including projector augmented-wave (GIPAW) method, which yielded relatively good results under perturbative external fields\cite{PhysRevB.63.245101,PhysRevB.76.024401}.

We developed a general first-principles calculation framework for calculating external magnetic fields in periodic systems\cite{74bb-qmp8}. This method utilizes a more flexible real-space lattice basis set, is applicable to calculating non-perturbative magnetic fields of arbitrary size, and supports calculations using general norm-conserving pseudopotentials. 
However, due to the relatively large size of the basis set and the difficulty in utilizing locality, this method introduces a significant computational burden, thus posing challenges in calculating larger (magnetic) units, i.e., smaller, experimentally obtainable magnetic fields.

In this paper, we implement a method for calculating finite magnetic fields in periodic systems based on the Gaussian basis. By constructing a specific magnetic Bloch basis, we introduce magnetic translational symmetry, thereby introducing the orbital effect of the magnetic field into the Kohn-Sham DFT calculation framework. 
Benefiting from the excellent locality properties of the Gaussian-type orbitals, this method effectively reduces the computational cost required for calculating applied magnetic fields and can simulate the behavior of solid materials under experimental magnetic fields. We demonstrate the method by calculating the Landau levels and zero-energy modes of monolayer graphene in laboratory-scale magnetic fields.
The entire algorithm is implemented in a developer's version of the Property Analysis and Simulation Package for Materials\cite{10.1063/5.0043703}.

\textit{\label{sec:level1}Methods}

We use atomic units throughout, setting $\hbar=m_e=e=1$, where $\hbar$ is the reduced Planck constant, e is the elementary
charge, and $m_e$ is the electron mass.
In the case of using separable pseudopotentials in the Kleinman-Bylander form\cite{PhysRevLett.48.1425}, 
the Hamiltonian of a system subject to a uniform external magnetic field $\mathbf{B}$ can be written as
\begin{equation}
\hat{H} = \frac{1}{2}\hat{\bm{\pi}}^2 + \hat{V}^{\text{GIPAW}} + V^{\text{loc}}(\mathbf{r})
\end{equation}
Here, \(\hat{\bm{\pi}}=-i\nabla+\mathbf A(\mathbf r)\) is the kinetic momentum, where $\mathbf{A}(\mathbf{r})$ is the magnetic vector potential. The lattice-periodic local potential \(V^{\mathrm{loc}}\) contains the local part of the pseudopotential and the electron-electron interaction potential. 
The nonlocal pseudopotential is written in separable Kleinman-Bylander form $\hat{V}^{\text{GIPAW}}=\sum_{\mu} \epsilon_{\mu} |\beta^B_\mu\rangle\langle \beta^B_\mu|$, 
with each projector carrying a magnetic phase that ensures commutation with magnetic translations\cite{PhysRevB.76.024401,PhysRevB.63.245101}.
\begin{equation}
|\beta^B_\mu \rangle = e^{-\mathrm{i}\mathbf{A}_{\text{Lan}}(\bm{\tau}_\mu)\cdot\mathbf{r}-\mathrm{i} f(\mathbf{r}-\mathbf{\bm{\tau}_\mu})} |\beta_\mu^0\rangle
\end{equation}

Here, $\mathbf{A}_{\text{Lan}}(\mathbf{r})$ represents the magnetic vector potential in the Landau gauge, and $f(\mathbf{r})$ is the gauge transformation factor, satisfying:
$\nabla f(\mathbf{r}) = \mathbf{A}_{\text{Lan}}(\mathbf{r})-\mathbf{A}_{\text{sym}}(\mathbf{r})$, where $\mathbf{A}_{\text{sym}}$ is the magnetic vector potential in the symmetric gauge. This magnetic phase transformation factor is significant because by 
defining the magnetic phase based on the Landau gauge can the magnetic Bloch basis satisfy the magnetic Bloch theorem, whereas atomic orbitals in the absence of a magnetic field can approximate those in a symmetric gauge magnetic field better\cite{PhysRevB.91.075122}.

We adopt this established GIPAW form without modification. The same gauge-including construction motivates the localized orbital basis introduced below.

Previous studies have demonstrated that when the magnetic flux satisfies the quantization condition—that is, when the magnetic flux is an integer multiple, $n_\Phi$, of $2\pi$, i.e. $\mathbf{B}\cdot(\mathbf{a}_1\times\mathbf{a}_2)=2\pi n_\Phi$ where $\mathbf{a}_1$ and $\mathbf{a}_2$ are primitive vectors—
the eigenstates $\{\psi_{i,\mathbf{k}}\}$ of Hamiltonian (Here, $i$ is the band index, and $\mathbf{k}$ lies within the first Brillouin zone) satisfy the magnetic Bloch theorem\cite{PhysRevLett.92.186402,LEE20071310}($\mathbf{R}$ represents the magnetic unit cell lattice vector):
\begin{equation}
\psi_{i,\mathbf{k}}(\mathbf{r}+\mathbf{R}) = e^{\mathrm{i}\mathbf{k}\cdot\mathbf{R}} e^{-\mathrm{i}\mathbf{A}(\mathbf{R})\cdot\mathbf{r}}
\psi_{i,\mathbf{k}}(\mathbf{r})
\end{equation}
This transformation law can be readily generalized to the case where the magnetic flux quantum number is an arbitrary rational number. When $n_\Phi = p/q$, the magnetic flux quantization condition can always be satisfied by expanding 
the chemical unit cell by a factor of $q$ into a larger ``magnetic unit cell'',
e.g. $\mathbf{a}_1^m=q\mathbf{a}_1^c,\mathbf{a}_2^m=\mathbf{a}_2^c$
(In the following derivation, lattice vectors refer to the magnetic lattice unless otherwise stated.).

In conventional LCAO calculations for periodic systems, crystalline eigenstates are expanded in Bloch sums of localized orbitals. In a magnetic field, the corresponding basis must instead transform according to the magnetic Bloch theorem.

We construct the following ``magnetic Bloch basis'' which satisfies the magnetic Bloch theorem, i.e. 
$b_{\mu\mathbf{k}}(\mathbf{r}+\mathbf{P}) = e^{\mathrm{i}\mathbf{k}\cdot\mathbf{P}} e^{-\mathrm{i}\mathbf{A}(\mathbf{P})\cdot\mathbf{r}}b_{\mu,\mathbf{k}}(\mathbf{r})$ ,
when the magnetic vector potential satisfies $\mathbf{A}(\mathbf{P})\cdot\mathbf{Q}=2\pi n$, in which $\mathbf{P}$,$\mathbf{Q}$ are arbitrary magnetic lattice vectors and $n$ is an integer. 
The subscript $\mu$ labels the atomic orbitals within the (magnetic) unit cell.
\begin{equation}
b_{\mu\mathbf{k}}(\mathbf{r}) = \frac{1}{\sqrt{N}} \sum_{\mathbf{R}} e^{\mathrm{i}\mathbf{k}\cdot\mathbf{R}} e^{\mathrm{i}\mathbf{A}(\bm{\tau}_\mu)\cdot\mathbf{R}} \phi^B_{\mathbf{R}\mu}(\mathbf{r}) 
\end{equation}
The magnetic atomic orbital basis $\phi_\mu^B$ can be expressed as a standard atom-centered orbital basis set multiplied by a magnetic phase factor resembling the GIPAW-style pseudopotential form
($\bm{\tau}_\mu$ represents the orbital center, i.e., the atomic position, within the magnetic unit cell). 
\begin{equation}
\phi^B_{\mathbf{R}\mu}(\mathbf{r}) = \mathrm{e}^{-\mathrm{i} \mathbf{A}_{\mathrm{Lan}}(\mathbf{R}+{\bm{\tau}}_\mu)\cdot\mathbf{r} 
- \mathrm{i} f(\mathbf{r}-\mathbf{R}-\bm{\tau}_\mu)} \chi_\mu(\mathbf{r}-\mathbf{R}-\bm{\tau}_\mu)
\end{equation}
Therefore, the wave functions expressed as linear expansions of these Bloch basis functions also satisfy the magnetic Bloch theorem.

\textit{\label{sec:level1}Implementation}

To facilitate the computation of the Hamiltonian and overlap matrix elements in real space of the following form, we use the Gaussian-type orbitals as a field-free basis set.
\begin{equation}
\begin{aligned}
S_{\mathbf{P}\mu;\mathbf{Q}\nu} &= \langle \phi^B_{\mathbf{P}\mu} | \phi^B_{\mathbf{Q}\nu}\rangle\\
H_{\mathbf{P}\mu;\mathbf{Q}\nu} &= \langle \phi^B_{\mathbf{P}\mu} | \hat{H} | \phi^B_{\mathbf{Q}\nu}\rangle = T_{\mathbf{P}\mu;\mathbf{Q}\nu} + V^{KB}_{\mathbf{P}\mu;\mathbf{Q}\nu}+V^{loc}_{\mathbf{P}\mu;\mathbf{Q}\nu}
\end{aligned}
\end{equation}
In our approach, we assume the magnetic field is along the $\mathbf{a}_3$ direction: $\mathbf{B}=\frac{2\pi n_\Phi}{\Omega}\mathbf{a}_3$, The magnetic vector potential under the Landau gauge can be defined as 
$\mathbf{A}_{\text{Lan}}(\mathbf{r}) = 2\pi n_\Phi (\mathbf{b}_1\cdot\mathbf{r})\mathbf{b}_2$, that under the symmetric gauge is $\mathbf{A}_{\text{sym}}=\frac{\pi n_\Phi}{\Omega}(\mathbf{a}_3\times\mathbf{r})$,
in which $\mathbf{a}_\alpha$ is the magnetic lattice vector and $\mathbf{b}_\beta$ satisfies $\mathbf{a}_\alpha \cdot \mathbf{b}_\beta=\delta_{\alpha\beta}$.
The magnetic phases modify the Gaussian basis functions but remain analytically evaluable of the overlap, kinetic, and nonlocal-projector integrals.
The formula derived earlier can be rearranged as follows, assuming the magnetic field direction is fixed (taking the magnetic field along the $\mathbf{a}_3$ direction as an example):
\begin{equation}
\begin{aligned}
\phi_{\mathbf{R}\mu}(\mathbf{r}) = e^{-\mathrm{i}\pi n_\Phi (\mathbf{b}_1 \cdot \mathbf{r})(\mathbf{b}_2 \cdot \mathbf{r})} e^{\mathrm{i}\theta_{\mathbf{R}\mu}}
&e^{-\mathrm{i}\mathbf{q}_{\mathbf{R}\mu}\cdot(\mathbf{r}-\mathbf{R}-\bm{\tau}_\mu)}\\
&\chi_\mu(\mathbf{r}-\mathbf{R}-\bm{\tau}_\mu)
\end{aligned}
\end{equation}

Here, the phase and shift vectors are defined as $\theta_{\mathbf{R}\mu} = -\pi n_\Phi (\mathbf{b}_1\cdot(\mathbf{R}+\bm{\tau}_\mu))(\mathbf{b}_2\cdot(\mathbf{R}+\bm{\tau}_\mu)),
\mathbf{q}_{\mathbf{R}\mu} = \pi n_\Phi ((\mathbf{b}_1\cdot(\mathbf{R}+\bm{\tau}_\mu))\mathbf{b}_2-(\mathbf{b}_2\cdot(\mathbf{R}+\bm{\tau}_\mu))\mathbf{b}_1)$. 

The plane-wave magnetic phase $e^{-\mathrm{i}\mathbf{q}_{\mathbf{R}\mu}\cdot(\mathbf{r}-\mathbf{R}-\bm{\tau}_\mu)} $can be incorporated as a complex shift of the Gaussian center. 
Therefore, various recursive methods can be used to analytically calculate the required component matrix elements. Overlap matrix elements is direct;
 nonlocal pseudopotential matrix elements are also overlapping matrix elements $\langle \beta_{\mathbf{P}\mu} | \phi_{\mathbf{Q}\nu}\rangle$. 
 Due to the rewriting of the momentum operator in the presence of the magnetic field, the form of the kinetic energy term changes somewhat, but all terms are simplified to 
 the same set of complex Gaussian overlapping integrals, derivative integrals, and position integrals (all of which can be analytically calculated through recursive relations\cite{10.1063/1.455717,10.1063/1.459751,10.1021/acs.jctc.7b00540,PhysRevE.64.056706}).
The matrix elements of the local potential can be calculated directly using real-space grid integrals from the standard LCAO method.

The matrix elements defined based on the magnetic orbital basis set no longer satisfy the general translation symmetry, but satisfy the following translation relationship.
\begin{equation}
X_{(\mathbf{P}+\mathbf{R})\mu,(\mathbf{Q}+\mathbf{R})\nu} = e^{-\mathrm{i}(\mathbf{A}({\bm{\tau}_\nu})-\mathbf{A}({\bm{\tau}_\mu}))\cdot\mathbf{R}} X_{\mathbf{P}\mu,\mathbf{Q}\nu}
\end{equation}
Therefore, $H(\mathbf{k})\equiv \langle b_{\mu}(\mathbf{k}) | \hat{H} | b_{\nu}(\mathbf{k}) \rangle$ and 
$S(\mathbf{k})\equiv \langle b_{\mu}(\mathbf{k}) | b_{\nu}(\mathbf{k}) \rangle$ in the secular equation $H(\mathbf{k})C_{i,\mathbf{k}}=\epsilon_{i,\mathbf{k}}S(\mathbf{k})C_{i,\mathbf{k}}$ (The component $c_{i,\mathbf{k},\mu}$ of $C_{i,\mathbf{k}}$ is the Bloch basis coefficient in the wavefunction)
 can be calculated using the following formula.
\begin{equation}
\begin{aligned}
X_{\mu\nu}(\mathbf{k}) = \sum_{\mathbf{R}} X_{\mathbf{0}\mu,\mathbf{R}\nu} e^{\mathrm{i}\mathbf{k}\cdot\mathbf{R}} e^{\mathrm{i}\mathbf{A}(\bm{\tau}_\mu)\cdot\mathbf{R}}
\end{aligned}
\end{equation}

For \(n_\Phi=p/q\), the magnetic unit cell contains \(q\) chemical unit cells, and hence the LCAO Hamiltonian dimension increases approximately linearly with \(q\).
Internally, the magnetic-cell orbital index can be resolved into a chemical-cell index and an intra-cell orbital index $\mu\to(\overline{\mathbf{R}}^c,\mu)$($\overline{\mathbf{R}}^c$ is the lattice vector corresponding to the chemical unit cell within a single magnetic unit cell).
Previous work has shown that the charge density exhibits strong periodicity, namely, the periodicity of the chemical unit cell rather than that of the magnetic unit cell\cite{74bb-qmp8}.
\begin{equation}
n({\mathbf{r}+\mathbf{R}^c}) = n(\mathbf{r})
\end{equation}
It can be shown that the matrix element blocks corresponding to different chemical unit cells satisfy certain translational relationships; utilizing these relationships avoids redundant matrix element calculations.

This reduction primarily avoids redundant Hamiltonian construction and does not reduce the dimension of the resulting magnetic eigenvalue problem. 
However, another important property is that, for certain systems, the difference in charge density between the magnetic-field and zero-field cases is much smaller than the corresponding differences in other physical properties, such as the band structure\cite{74bb-qmp8}. 

Therefore, when only a few states near the Fermi level are required in the weak-field regime, we can start from the zero-field charge density (defined within the zero-field unit cell) and perform a single-step, non-self-consistent calculation using low-scaling algorithms, 
such as Kernel Polynomial Method (KPM)\cite{KPM1,KPM2,KPM3}.

\textit{\label{sec:level1}Results}

In the calculations, we used separable pseudopotentials of the GTH form\cite{PhysRevB.54.1703,Krack2005GTH} and the corresponding Gaussian basis sets\cite{VandeVondele2007MOLOPT}.

Landau level of monolayer graphene system: We used this method to calculate the Landau levels of a monolayer graphene system. 
Theoretically, the Landau levels of monolayer graphene in a vertical magnetic field should satisfy the following formula\cite{gll1}:
\begin{equation}
E_{n} = \mathrm{sgn}(n) v_{F} \sqrt{2e\hbar|\mathbf{B}||n|}
\end{equation}

In the calculations, we used the zero-field charge density to perform non-self-consistent calculations undera magnetic field perpendicular to the graphene plane, using the LDA functional\cite{PhysRevLett.45.566,PhysRevB.45.13244} and ignoring spin and Zeeman effect. Given that diagonalizing the full Hamiltonian entails a massive computational
load—whereas calculating Landau levels requires information only in the vicinity of the Fermi level—we employed the KPM\cite{KPM2} to estimate the Fermi level and use a Krylov subspace method
to solve for the 100 energy levels surrounding it; the resulting data show good agreement with theoretical formulas in the low-magnetic-field regime,
and deviations emerge for higher Landau indices and stronger fields, consistent with the breakdown of the low-energy Dirac approximation\cite{PhysRevB.106.155414,YUAN20121446}.
At the same time, our calculation results confirm another important property of graphene in a magnetic field: the existence of zero-energy modes\cite{PhysRevA.19.2461,PhysRevLett.95.146801}. 
Moreover, by visualizing the two sets of wave functions at the Fermi level, we found that the two zero-energy states are sublattice polarized: one is supported predominantly on the A sublattice and the other on the B sublattice
\begin{figure*}[h]
  \includegraphics[width=\textwidth]{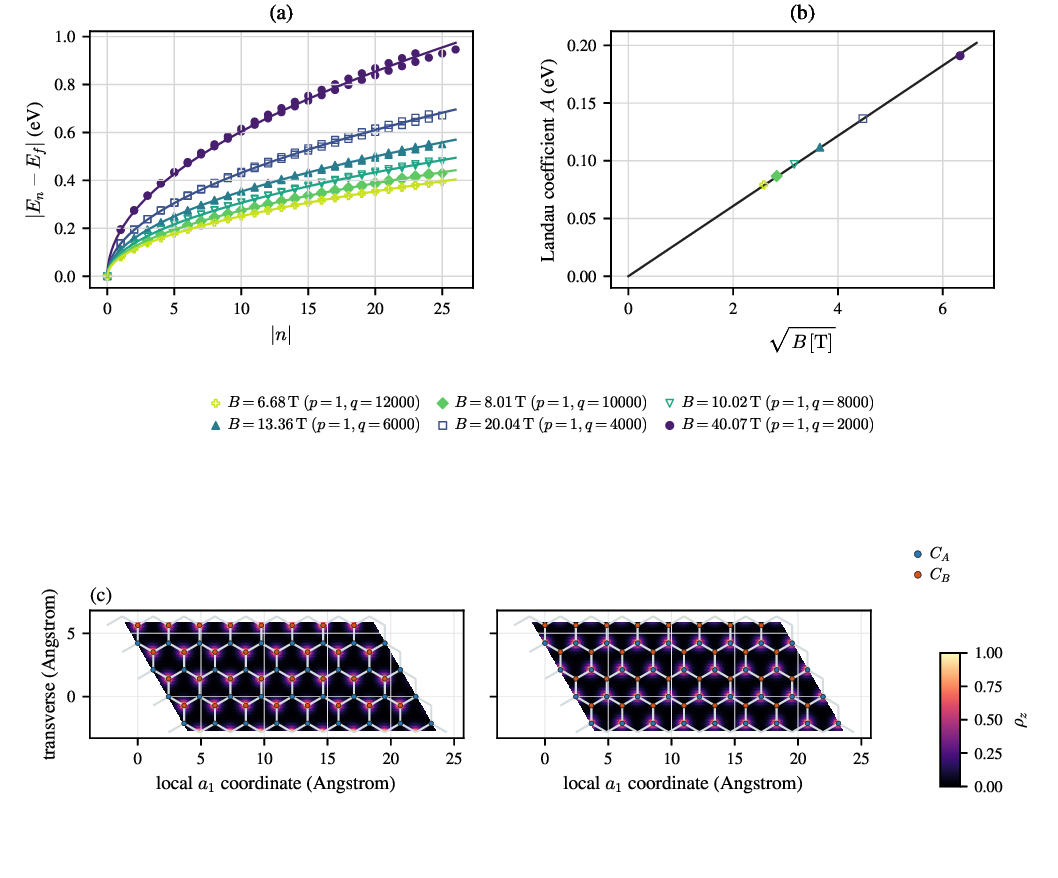}
  \caption{\label{fig:epsart} (a) Landau levels under different magnetic field strengths. The scatter plot represents the actual calculated energy levels, and the solid line represents the result of fitting the theoretical formula.
  Neglecting spin, the approximate doubly degenerate derives from $\mathbf{K}$ and $\mathbf{K}'$.
  (b) Fitting coefficients of Landau levels and $\sqrt{|\mathbf{B}|}$ ($A$ is the $ v_{F} \sqrt{2e\hbar|\mathbf{B}| } $ in the theoretical formula)  
  (c) The two zero-energy modes corresponding to $p=1,q=10,000$(The corresponding magnetic field strength is approximately 8.0152 T.).A portion of the crystal lattice was captured; $C_A$ and $C_B$ are two non-equivalent carbon atoms in graphene.
  the image shows the distribution of the wavefunction modulus after integration along the z-direction, i.e. $\int |\psi({\mathbf{r}})|^2 \mathrm{d}z$. }
\end{figure*} 

\textit{\label{sec:level1}Conclusion}

We have developed a density functional theory (DFT) program for applied magnetic fields based on the Gaussian-type orbitals. Thanks to the LCAO method, this program effectively reduces the computational cost of applied magnetic field DFT and can support self-consistent calculations for larger unit cells and smaller magnetic fields; 
for systems with small chemical unit cells, it can perform electronic structure calculations under experimental magnetic fields.

\bibliography{reference}

\end{document}